\documentclass[fleqn,usenatbib]{mnras}

\usepackage{newtxtext,newtxmath}
\usepackage[T1]{fontenc}
\usepackage{ae,aecompl}

\DeclareRobustCommand{\VAN}[3]{#2}
\let\VANthebibliography\thebibliography
\def\thebibliography{\DeclareRobustCommand{\VAN}[3]{##3}\VANthebibliography}

\usepackage{graphicx}	% Including figure files
\usepackage{amsmath}	% Advanced maths commands

\newcommand{\Gaia}{\textit{Gaia} }

\newcommand{\tilvrot}{$\widetilde{V_{\phi}}$ }

\newcommand{\lv}{$l_v$ }
\newcommand{\ltsim}{\protect\raisebox{-0.5ex}{$\:\stackrel{\textstyle <}{\sim}\:$}}

\title[Growing and disrupting two neighbouring spiral arms]{Clues to growth and disruption of two neighbouring spiral arms of the Milky Way}

\author[N. Funakoshi et al.]{
Natsuki Funakoshi,$^{1}$\thanks{E-mail: \href{mailto:n.funakoshi@ucl.ac.uk}{n.funakoshi@ucl.ac.uk}}%
Noriyuki Matsunaga,$^{2}$
Daisuke Kawata,$^{1}$
Junichi Baba,$^{3,4}$
Daisuke Taniguchi$^{4}$
\newauthor{
and Michiko Fujii$^{2}$
}
\\
$^{1}$Space and Climate Physics, Mullard Space Science Laboratory, University College London Holmbury St. Mary, Dorking, Surrey, RH56NT, UK\\
$^{2}$The University of Tokyo, 7-3-1, Hongo, Bunkyo-ku, 113-0033, Tokyo, Japan\\
$^{3}$Kagoshima University, 1-21-35, Korimoto, Kagoshima 890-0065, Japan\\
$^{4}$National Observatory of Japan, Mitaka-shi, Tokyo, 181-8588, Japan
}

\date{Accepted XXX. Received YYY; in original form ZZZ}

\pubyear{2024}

\begin{document}
% \label{firstpage}
% \pagerange{\pageref{firstpage}--\pageref{lastpage}}
\maketitle

\begin{abstract}
Studying the nature of spiral arms is essential for understanding the formation of the intricate disc structure of the Milky Way. The European Space Agency's \Gaia mission has provided revolutionary observational data that have uncovered detailed kinematical features of stars in the Milky Way. However, so far the nature of spiral arms continues to remain a mystery.
Here we present that the stellar kinematics traced by the classical Cepheids around the Perseus and Outer spiral arms in the Milky Way shows strikingly different kinematical properties from each other: the radial and azimuthal velocities of Cepheids show positive and negative correlations in the Perseus and Outer arms, respectively. We also found that the dynamic spiral arms commonly seen in an N-body/hydrodynamics simulation of a Milky Way-like galaxy can naturally explain the observed kinematic trends. 
Furthermore, a comparison with such a simulation suggests that the Perseus arm is being disrupted while the Outer arm is growing.
Our findings suggest that two neighbouring spiral arms in distinct evolutionary phases - growing and disrupting phases - coexist in the Milky Way.
\end{abstract}
\begin{keywords}
Galaxy: disc --- Galaxy: kinematics and dynamics --- Galaxy: structure --- stars: variables: Cepheids --- galaxies: spiral.
\end{keywords}

\section{Introduction}\label{sec1}

Spiral arms are one of the major structures of disc galaxies, where a majority of stars form. Exploring the nature of spiral arms is crucial for understanding the structure formation of disc galaxies. Previous theoretical studies have proposed two major scenarios to explain the nature of spiral arms, both of which remain in debate \citep{Dobbs2014}. The density wave scenario, including kinetic density wave \citep{Kalnajs1973}, proposes rigidly rotating, long-lived spiral arms \citep{Lin1964}. In this scenario, spiral arms rotate at a constant pattern speed irrespective of the galactocentric radius, while the azimuthal velocity of stars changes as a function of the galactocentric radius. Another major scenario is the dynamic arm scenario \citep{Baba2013,Baba2018}. In this scenario, spiral arms rotate at a similar speed to the stars, exhibiting transient and recurrent patterns as seen in many N-body simulations \citep[e.g.][]{Fujii2011,Grand2012,Grand2013,Baba2013,Kawata2014} and analytical studies \citep[e.g.][]{Meidt2023}. The transient and recurrent spiral arms can be explained by the superposition of multiple transient patterns \citep{Roskar2012} or different spiral arm modes with different pattern speeds \citep{Minchev2006,Quillen2011,Sellwood2014,Sellwood2019,Sellwood2021}. Some studies further suggest the nonlinear growth also enhances the spiral arm features when the modes overlap \citep[e.g.][]{Kumamoto2016}.

One way of studying the nature of spiral arms through observations is to test whether there is any offset in spatial positions of spiral arms traced by young and old objects. According to the density wave scenario, spiral arms traced by objects of different ages are offset with respect to each other, and the offset angle depends on the galactocentric radius \citep{Dobbs2010,Baba2015}. In contrast, the dynamic arm scenario predicts no offset \citep{Grand2012,Baba2015}. 
However, there is a mixture of observational indications supporting either the density wave scenario for the Milky Way \citep[e.g.][]{Hou2015} and for external galaxies \citep[e.g.][]{Abdeen2022,Vallée2020} or the dynamic arm scenario for the Milky Way \citep[e.g.][]{Castro-Ginard2021} and for external galaxies \citep[e.g.][]{Foyle2011,Ferreras2012}.

An alternative approach is to look in detail at the distribution of stellar kinematics around spiral arms. Vertex deviation is a convenient concept to characterise stellar kinematics. Vertex deviation is angle of inclination of velocity ellipsoid between galactocentric radial motion and azimuthal motion of stars.
Previous theoretical studies suggest that the density wave scenario creates spiral arms with different signs of vertex deviation depending on which side of the spiral arm is analysed \citep{Roca-Fabrega2014}. On the contrary, the sign of vertex deviation of the dynamic spiral arms does not necessarily change around the peak of the spiral arm. 
In a previous study \citep{Baba2018}, vertex deviation around the Perseus arm in the Milky Way was calculated by analysing the kinematics of classical Cepheid variables. Although they suffered from a small number statistics, they claimed that there was no change in vertex deviation between in front of and behind a peak of the spiral arm, i.e. the expected change in the sign of vertex deviation around a peak of the spiral arms in the density wave scenario was not observed. However, for a comprehensive and unambiguous understanding of the nature of spiral arms, it is necessary to reveal kinematics of multiple spiral arms.
This calls for analysis of stellar kinematics to be studied over a wider region of the Galactic disc, covering both sides of multiple spiral arms. To this end, it is crucial to have tracer stars with precisely measured distances to distinguish whether they are located at leading or trailing sides of spiral arms. Additionally, a sufficient number of stellar samples is required to unambiguously measure whether there are any changes in a sign of vertex deviation.

The Milky Way has a bar structure in the centre of its disc. Hence, the spiral arms in the inner disc could be more affected by the Galactic bar. Therefore, by focusing on the kinematics of the spiral arms located in the outer region of the Galactic disc, i.e. Perseus and the Outer arms, we can investigate simpler kinematics of the spiral arm. 
Hence, we focus on comparing stellar kinematics of these two spiral arms, which would be crucial for understanding the nature of spiral arms.

Investigating kinematics of the Perseus and Outer arms requires precise mapping of stars' positions, even further away than 4 kpc from the Sun \citep{Reid2019}.
The European Space Agency's \Gaia mission has provided revolutionary observational data that are invaluable for studying kinematic structures of the Galactic disc by measuring distances and motions of an unprecedentedly large number of stars with remarkable precision. However, the distance uncertainties arising from their parallax measurement errors become discernibly large when stars are at distances greater than about 4~kpc with the current measurement of \textit{Gaia}'s third data release \citep{Vallenari2023}. \textit{Gaia}'s current distance measurements are not precise enough to tell on which side of the spiral arm stars around the Outer arm are located.

Classical Cepheids, which are pulsating variable stars with masses between 4 and 20 M$_{\sun}$, can be used to obtain highly precise distance measurements even as far away as distance to the Outer arm. The uncertainties in distances obtained from the period-luminosity relation are less than 4\% for about 90\% of the Cepheids found in the Milky Way \citep{Skowron2018,Skowron2019}, which surpasses that of \textit{Gaia}'s parallax distance. This level of precision enables the identification of whether stars around the Outer arm are on the leading or trailing side of the spiral arm. We cross-match the Cepheid data with the \Gaia DR3, to obtain six dimensions of phase space information of stars up to around the Outer arm. The combination of these recently available superb data enables us to study stellar kinematics around both the Perseus and Outer arms.

The aim of this paper is to probe the nature of the spiral arm from the stellar kinematics around the Perseus and Outer arms. In Section \ref{sec:Method}, we describe the observational data and our approach to measure stellar kinematics. Section \ref{sec:motion} presents the stellar kinematics shown by the Cepheids around the Perseus and Outer arms. In Section \ref{sec:compDYN}, we show the kinematic structure of the spiral arms in an N-body simulation of a disc galaxy and compare it with the observed spiral arms to provide a new picture of the nature of the spiral arms and their formation. In Section \ref{sec:summary}, we discuss and summarise our findings.

\section{Methods}\label{sec:Method}
\subsection{Observational data} \label{sec:sample}

Our sample was taken from a catalogue of classical Cepheids \citep{Skowron2018,Skowron2019}.
A majority of these Cepheids were identified by the Optical Gravitational Lensing Experiment (OGLE), which explores the Galactic disc up to approximately 20 kpc from the Galactic centre. 
Distances of the Galactic classical Cepheids were estimated from a period-luminosity relation \citep{Wang2018} and an extinction map \citep{Bovy2016}. 

We selected the data with a small distance uncertainty of $\sigma_D/D<0.05$, where $D$ and $\sigma_D$ refer to the period-luminosity distance from the Sun and its uncertainty, respectively. $\sigma_D$ is not the uncertainty in distance modulus, but rather the uncertainty in distance propagated from the uncertainty in distance modulus \citep{Skowron2018}.
The Cepheids data were cross-matched with the \Gaia DR3 sources \citep{GaiaCollaboration2022} and combined with their proper motion and line-of-sight velocity. We applied a quality cut on the \Gaia data, selecting only the data with $\tt{RUWE}$$< 1.4$, $\tt{non\_single\_star}$$ = 0$ and $\tt{radial\_velocity\_error}<$10~$\mathrm{km~s^{-1}}$.

To obtain information on the position and velocity of the Cepheid in the Galactic rest frame, we utilised solar positions and motions which were obtained in a previous research using High Mass Star Forming Regions \citep[HMSFRs,][]{Reid2019}. The solar positions applied are the Galactocentric radius of $R_0 =8.15$~kpc and distance from the Galactic mid-plane of $Z_\odot=0.0055$~kpc. Solar motions are defined as velocities $(V_{R,\odot},V_c+V_{\phi,\odot},V_{Z,\odot})=(-10.6,246.7,7.6)$~$\mathrm{km~s^{-1}}$, where $V_{R,\odot}$, $V_{\phi,\odot}$ and $V_{Z,\odot}$ are solar proper motion in the direction to the Galactic anti-centre, the direction of the Galactic rotation and the direction towards the north Galactic pole, respectively.

From these conditions, we obtained the position, $(R, \phi, Z)$, and the velocity, $(V_R, V_{\phi}, V_Z)$, for each Cepheid in the Galactocentric cylindrical coordinate, where $R$ and $Z$ are the Galactocentric radius and the distance from the true Galactic mid-plane without warp, respectively. $V_R$, $V_{\phi}$, and $V_Z$ denote velocity radially outward from the Galactic centre, azimuthal velocity, and vertical velocity, respectively.

From the six-dimensional data (3D position and 3D velocity) obtained, we only included the data that satisfy following criteria to exclude the data that are improbable to be classical Cepheids :
\begin{eqnarray}
&R<20 \, \mathrm{kpc},\label{eq:1}\\ 
&|Z|<3 \, \mathrm{kpc} ,\\
    &|V_Z|<50\, \mathrm{km~s^{-1}},\\
    &|V_R|<50\, \mathrm{km~s^{-1}},\\
    &170\, \mathrm{km~s^{-1}}<V_{\phi}<270\, \mathrm{km~s^{-1}}\label{eq:5}.
\end{eqnarray}
Although Cepheids show warp and flare structures around the outer edge of the Galactic disc, no Cepheids are expected at $|Z|>3$~kpc \citep{Skowron2018,Skowron2019,Lemasle2022}.
The distances to Cepheids whose derived positions are at $R>20$~kpc or $|Z|>3$~kpc are likely to be overestimated compared to their actual distance, likely due to misclassifications with Type II Cepheids or binary stars.

Additionally, we limited the dataset to Cepheids with \textit{ecc} $<0.2$, where \textit{ecc} is the orbital eccentricity of the star computed in the \texttt{MilkyWayPotential2022} potential using \texttt{Gala} package \citep{2017JOSS....2..388P}. 
We scaled the potential to match the solar position to ($R_0$, 0, $Z_{\sun}$) as defined above and its circular velocity $V_c(R_0)$ to 228.9 km s$^{-1}$, which is derived from the mean rotation velocity of our Cepheids whose guiding radius, $R_g$, is around $R_0$. We consider that the mean rotation velocity for Cepheids is close to the circular velocity because they are kinematically cold \citep[e.g.,][]{Kawata2018,Almannaei2024} and their asymmetric drift is negligible.

 We obtained 1296 Cepheids, which cover $4<R<20$ kpc and $|\phi|<100^\circ$, where the Galactocentric angle, $\phi$, is defined as $0^\circ$ on the line between the Galactic centre and the Sun and positive in the direction of Galactic rotation.
 Fig.~\ref{fig:xy} shows the distribution of Cepheids in the XY coordinates. The analysis in this paper was limited to objects within the range of $-50^\circ<\phi<30^\circ$, since these regions encompass a significant number of Cepheids located around the position of the Perseus and Outer arms. 

The position of the Perseus arm was taken from a previous study \citep{Hou2021}. The position of the Outer arm was shifted radially outwards by 300 pc with respect to the position of the Outer arm reported in a previous study with HMSFRs \citep{Reid2019}, to align it with the position where the Cepheids are more concentrated, but keeping the pitch angle same as what is obtained with the HMSFRs. This does not necessarily imply that there is an offset in the positions of the spiral arm formed by HMSFRs and Cepheids. The previous study reported only 11 HMSFRs around the Outer arm, while 110 Cephids are found around the location of 300 pc outside of the position with the HMSFRs. Thus, we consider that the Cepheids' concentration is more likely to represent the real Outer arm position than the one suggested by a limited number of the HMSFRs reported before. 

\subsection{Calculation for Vertex Deviation}\label{sec:calc_lv}
The rotational velocity with respect to circular velocity, $\widetilde{V_{\phi}}$, was computed by subtracting circular velocity, $V_c$, from azimuthal velocity, $V_{\phi}$. 
We derived a circular velocity of the $n$-th Cepheids with angular momentum, $L_{z,n}$, by taking an average of the rotation velocity of Cepheids within $|L_{z}-L_{z,n}|<200$ kpc km s$^{-1}$.
Fig. \ref{fig:rotation} shows the derived circular velocity as a function of the guiding radius, $R_g$.

Vertex deviation refers to the tilt of the major axis of the velocity ellipsoid. The general form of the equations used to calculate the vertex deviation is given as follows \citep{Vorobyov2008}:
\begin{eqnarray}
l_v=\left\{
\begin{array}{ll}\nonumber
\displaystyle \frac{1}{2}\arctan \left( \frac{2\sigma^2_{R\phi}}{\sigma_{RR}^2-\sigma_{\phi\phi}^2}\right), & \quad(\sigma^2_{RR}>\sigma^2_{\phi\phi})\\
\displaystyle \frac{1}{2}\arctan \left( \frac{2\sigma^2_{R\phi}}{\sigma_{RR}^2-\sigma_{\phi\phi}^2}\right)+\mbox{sign}\left( \sigma^2_{R\phi} \right) \frac{\pi}{2}, &\quad (\sigma^2_{RR}<\sigma^2_{\phi\phi}),
\end{array}
\right.
\end{eqnarray}
where $\sigma_{ij}$ is the general velocity dispersion tensor
\begin{eqnarray}
\sigma_{ij}^2\equiv\overline{(v_i-\bar{v_i})(v_j-\bar{v_j})},
\end{eqnarray}
where $i$ and $j$ denote coordinate directions. Here the velocity dispersion tensor was computed using $V_R$ and $\widetilde{V_{\phi}}$, and for example, $\sigma^2_{R\phi} = \overline{(V_R-\overline{V_R})(\widetilde{V_{\phi}}-\overline{\widetilde{V_{\phi}}})}$.
This definition of \lv differs from the conventional definition of $l_v$. In the conventional definition, $V_{\phi}$ is used instead of $\widetilde{V_{\phi}}$. We combined the data around each spiral arm at a large range of the radius and hence $L_z$. It is therefore better to take into account the different circular velocity at the different $L_z$.
In addition, positive angles in the conventional definition of $l_v$ indicate a negative correlation between $V_R$ and $V_{\phi}$ \citep{Baba2018,Vorobyov2008,Roca-Fabrega2014}.
On the other hand, we defined the positive angle of \lv as a positive correlation between $V_R$ and $\widetilde{V_{\phi}}$. We consider that this is more natural definition than the conventional one, and we used our own definition of the sign of vertex deviation. 

The uncertainties in the vertex deviation were computed using a jack-knife method. Notably, the error estimation using the jack-knife method does not account for observational uncertainties of the Cepheids.
However, we found that the errors obtained by a jack-knife method are larger than the statistical errors computed by Monte Carlo sampling from the observational uncertainties in 
period-luminosity distances, \Gaia DR3 line-of-sight velocity and proper motion. This means that the Poisson noise has a greater impact than the uncertainty of each data point. Therefore, we employed the uncertainties estimated with the jack-knife method, to be conservative. \\

\begin{figure*}
% \epsscale{1.2}
\centering
\includegraphics[width=0.6\linewidth]{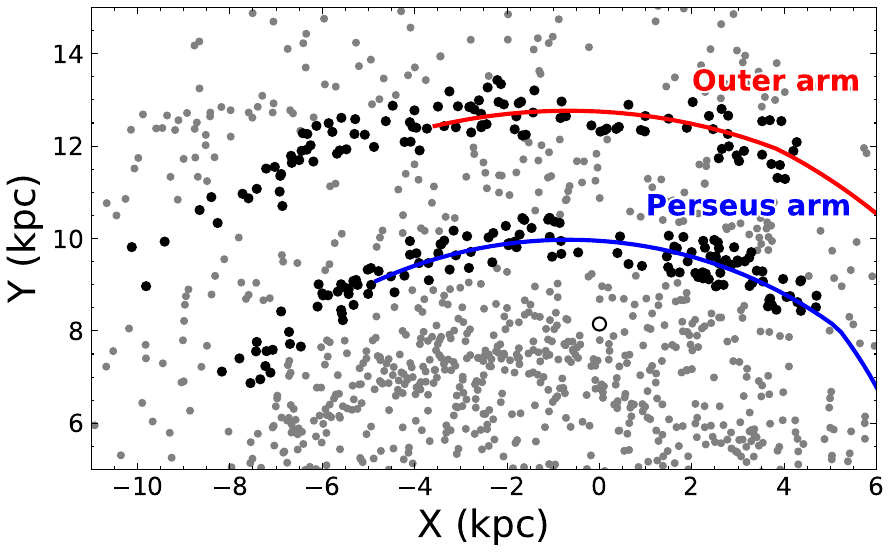}
\caption{The distribution of classical Cepheids obtained after cross-matching the classical Cepheids catalogue \citep{Skowron2018,Skowron2019} and the \Gaia DR3.
The black dots are the Cepheids around the spiral arm used in Fig.~\ref{fig:vrvphi}, and the grey dots are the other Cepheids. The black open circle shows the position of the Sun.
The purple and red lines indicate the Perseus and Outer arms respectively, as suggested in the literature \citep{Hou2021,Reid2019}. 
  \label{fig:xy}
  }
\end{figure*}

\begin{figure*}
% \epsscale{1.2}
\centering
\includegraphics[width=0.6\linewidth]{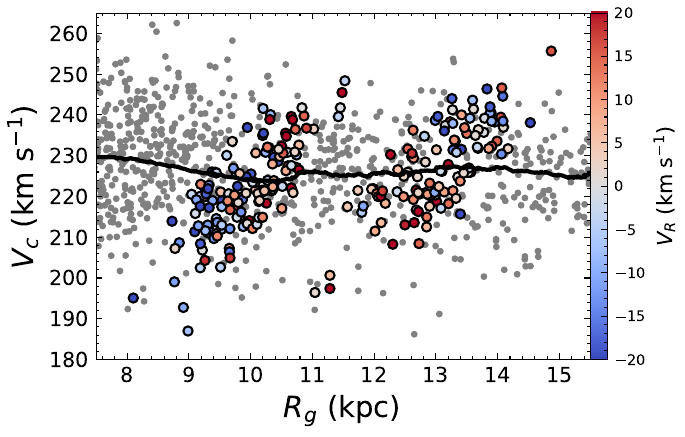}
\caption{The $R_g-V_c$ diagram with rotation curve derived from Cepheids.
The solid black line represents the circular velocity computed using Cepheids as mentioned in Section \ref{sec:calc_lv}. The dots with colour show Cepheids around the Perseus and Outer arm (same ones highlighted with black dots in Fig.~\ref{fig:xy}) and they are colour-coded with radial velocities, while grey dots show the rest of Cepheids.
  \label{fig:rotation}
  }
\end{figure*}

We analyse the vertex deviation by dividing the Cepheid samples around the Perseus and Outer arms into four groups based on their positions on the leading and trailing sides of the arms. Fig.~\ref{fig:philnR} illustrates the defined regions for groups (a) and (b) around the Outer arm (leading and trailing sides, respectively), and groups (c) and (d) around the Perseus arm (leading and trailing sides, respectively). Fig.~\ref{fig:philnR} displays a logarithmic radius on its vertical axis to trace logarithmic shapes of the spiral arms. 
The width of areas indicated as (a)-(d) is defined by $R' \equiv \ln R-\phi\times \tan 4^\circ$ to take into account the pitch angles of the spiral arms, which are about 4 deg for both the Perseus and Outer arms \citep{Reid2019,Hou2021}. These regions are in the range of (a) $2.55<R'<2.6$, (b) $2.5<R'<2.55$, (c) $2.3<R'<2.35$ and (d) $2.25<R'<2.3$, within $-50^\circ<\phi<30^\circ$. Cepheids belonging to any of these four groups are highlighted with black dots in Fig.~\ref{fig:xy}.

\section{Results}\label{sec:motion}

\begin{figure*}
% \epsscale{1.2}
\centering
\includegraphics[width=0.4\linewidth]{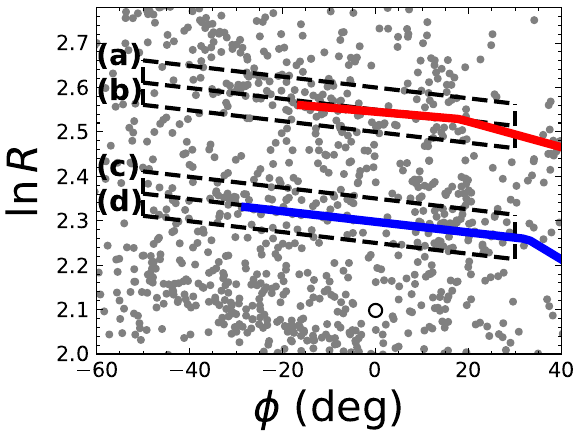}
\caption{Spatial distribution of Cepheids (grey dots), highlighting the areas selected to sample the Cepheids on the leading and trailing sides of the Outer and Perseus arms. 
The black open circle and the purple and red lines indicate the position of the Sun and two spiral arms as in Fig.~\ref{fig:xy}. 
  \label{fig:philnR}
  }
\end{figure*}

\begin{figure*}
% \epsscale{1.2}
\centering
\includegraphics[width=0.8\linewidth]{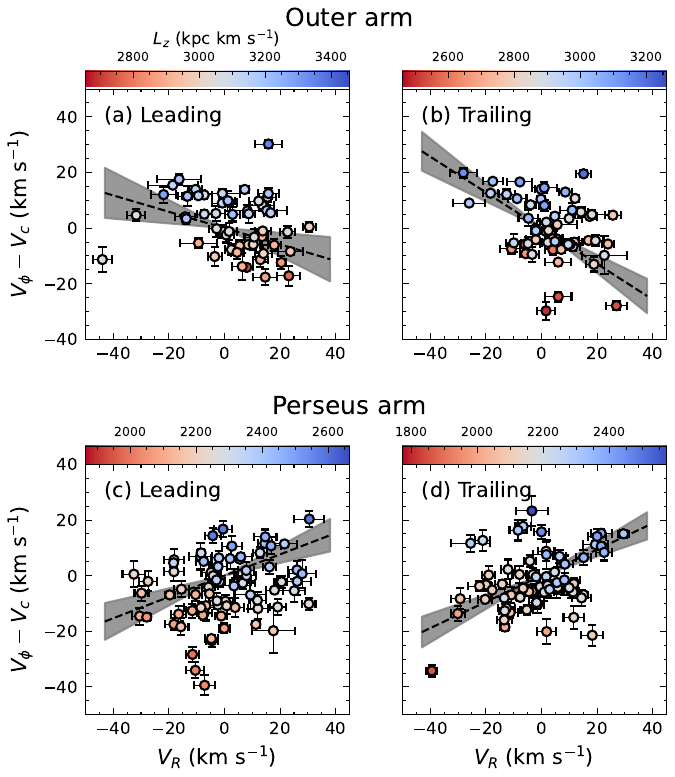}
\caption{The distribution between $V_R$ and $\widetilde{{V_{\phi}}}\equiv V_{\phi}-V_c$ of Cepheids in boxes highlighted as (a), (b), (c) and (d) in Fig.~\ref{fig:philnR}, respectively.
The colours represent the angular momentum $L_z = R \times V_{\phi}$. The dashed lines and grey shadows in each panel indicate the measured vertex deviation and its uncertainty.
  \label{fig:vrvphi}
  }
\end{figure*}

Fig.~\ref{fig:vrvphi} shows the correlation between $V_R$ and $\widetilde{V_{\phi}}=V_{\phi}-V_c$ for the Cepheids in the four groups around the Perseus and Outer arms.
Panels (a) and (b) in Fig.~\ref{fig:vrvphi} show results of Cepheids in the (a) trailing and (b) leading regions of the Outer arm, as indicated in Fig. \ref{fig:philnR}, respectively.
Panels (c) and (d) in Fig. \ref{fig:vrvphi} present results of Cepheids in the (c) leading and (d) trailing regions of the Perseus arm, as shown in Fig. \ref{fig:philnR}.
Angular momentum, $L_z$, is depicted by the colours of dots. 
The angle of the inclination of the velocity distribution is defined as vertex deviation, $l_v$. As described in Section \ref{sec:calc_lv}, the dashed line and the shaded area in each panel of Fig.~\ref{fig:philnR} show the slope of $l_v$, and its uncertainty computed with Cepheids in each region, respectively. Regions (a), (b), (c) and (d) contain 54, 56, 74 and 64 Cepheids, respectively.

Cepheids around the Perseus arm show positive correlations between $V_R$ and \tilvrot on both the leading and trailing sides. We obtain $l_v=0.38\pm0.15$ in region (c) and $l_v=0.47\pm0.13$ in region (d), respectively. Cepheids around the Outer arm have negative correlations on both sides, with $l_v\sim-0.29\pm0.21$ on the leading side, region (a), and $l_v\sim-0.64\pm0.16$ on the trailing side, region (b). 
The key point is that the vertex deviations are positive on both sides of the Perseus arm, while they are negative on both sides of the Outer arm. The vertex deviations around the Perseus and Outer arms differ significantly from each other, by more than 3 sigma.

\begin{figure*}
% \epsscale{0.6}
\centering
\includegraphics[width=0.6\linewidth]{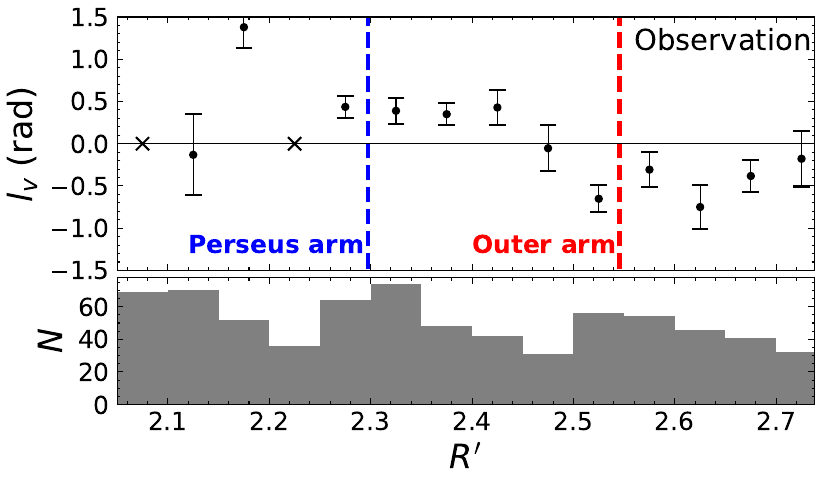}
\caption{Upper: The vertex deviation computed using $V_R$ and \tilvrot of our Cepheids as a function of $R'=\ln R-\phi\times \tan 4^\circ$.
Black solid circles with the error bars indicate the vertex deviation and the uncertainty in each $R'$ bin. Crosses indicate that the uncertainties in these $R'$ bins are too large to obtain meaningful vertex deviations.
Lower: Number distribution of Cepheids as a function of $R'$. The vertical purple and red dashed lines in the upper panel indicate the location of the Perseus and Outer arm, respectively. The Cepheids show positive vertex deviations around the Perseus arm and negative vertex deviations around the Outer arm. \label{fig:R-VD}}
\end{figure*}

Fig.~\ref{fig:R-VD} (upper panel) shows vertex deviation calculated as a function of $R'$. The lower panel of Fig.~\ref{fig:R-VD} shows a number of Cepheids in each $R'$ bin. The vertical purple and red dashed lines highlight the positions of the Perseus and Outer arms, respectively. Excesses of the number of Cepheids are seen around these vertical lines, which confirms the stellar density excess in the location of the spiral arm. As illustrated in Fig.~\ref{fig:vrvphi}, \lv are positive on both sides of the Perseus arm, while they are negative around the Outer arm.
It is noteworthy that the kinematic trends around these two neighbouring spiral arms are completely opposite.

\section{Discussion}\label{sec:compDYN}
\subsection{Classical density wave scenario}
Vertex deviation around the spiral arms can be derived analytically under the density wave scenario with tight-winding approximation \citep{Roca-Fabrega2014}. According to the density wave scenario, sign of \lv should change between trailing and leading sides of spiral arms. However, our observational data show $l_v>0$ on both sides of the Perseus arm and $l_v<0$ on both sides of the Outer arm as shown in Figs \ref{fig:vrvphi} and \ref{fig:R-VD}. This does not align with the expected behaviour of \lv in the density wave scenario.

\subsection{Comparison with spiral arms in N-body simulation}

The theoretical prediction from the density wave scenario does not explain well the observed kinematics structure of the Cepheids shown in Fig.~\ref{fig:vrvphi}.
On the other hand, we discovered that the trend of the opposite vertex deviation of two neighbouring spiral arms is naturally seen in a typical N-body/hydrodynamics simulation. This kind of simulation models commonly have spiral arms that follow the dynamic arm scenario. 

We used an N-body/hydrodynamics simulation of a Milky Way-like isolated disc galaxy \citep{Baba2020,Baba2021}, which includes self-gravity, radiative cooling, star formation and stellar feedback \citep{Saitoh2008,Saitoh2009,Saitoh2010,Baba2017}.
The simulation also self-consistently includes stellar and gaseous discs, a classical bulge and a fixed dark matter halo, with particle masses of the star and gas approximately $9.1 \times 10^3 M_\odot$ and $3\times 10^3 M_\odot$, respectively. The gravitational softening length is 10 pc, which is sufficiently small to resolve the three-dimensional structure of the disc galaxy \citep{Baba2013}. The analysis of the simulation data for this study was limited to the particles of approximately same age as the Cepheids. The ages of our sampled Cepheids estimated from the period-age relation \citep{Anderson2016a} were taken from the catalogue, where their mentality is considered to follow the observed metallicity gradient \citep{Genovali2014} and the metallicity dependence of the period-age relation is taken into account.
We then obtained a well-known dependence of ages of Cepheids on a Galactocentric radius, $R$, which is attributed to the metallicity gradient. Based on this relation, we applied a selection of particles with $6.67\times (R/\mathrm{kpc})$\ltsim age/Myr\ltsim $26.7\times (R/\mathrm{kpc})$. 
% \textbf{}

\subsubsection{Neighbouring spiral arms with different kinematics in N-body simulation}\label{sec:compneigh}

\begin{figure*}
% \epsscale{0.6}
\centering
\includegraphics[width=0.5\linewidth]{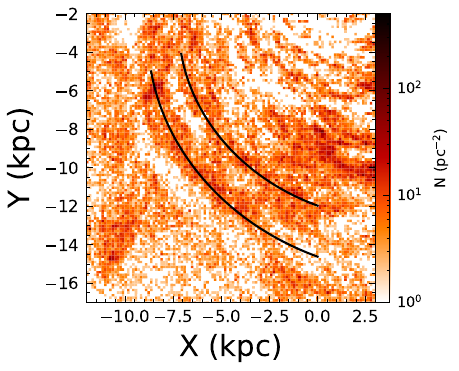}
\caption{The dentisy distribution of the Cepheid-like star particles at 3.85 Gyr in the N-body/hydrodynamics simulation of a Milky Way-like isolated disc galaxy. The neighbouring two spiral arms indicated with solid black lines are the spiral arms used in Fig.~\ref{fig:r-lv_2}. The vertical dashed line shows a line defined as $\phi = 0^\circ$.
  \label{fig:snap_3581}}
\end{figure*}

Fig.~\ref{fig:snap_3581} illustrates a snapshot of two neighbouring spiral arms, which show similar kinematical trends to our observed Cepheids.
The dashed line in Fig.~\ref{fig:snap_3581} shows a line defined as $\phi=0^\circ$ and $\phi$ is defined as positive in clockwise direction (the galactic rotation direction).
These arms were found at 3.58 Gyr in the simulation time.
Using the particle data within $0^\circ<\phi<60^\circ$, vertex deviation was computed as a function of $R'' \equiv \ln R - \phi  \times \tan 20^\circ$ which are approximately perpendicular to both of the two spiral arms. The $V_c$ used to compute vertex deviation was derived in the same manner as described in Section \ref{sec:calc_lv}.
Note that the spiral arms have a pitch angle of around 20 deg, which is significantly larger than those of the Perseus and Outer arms. However, our focus is on vertex deviation, and we accept this discrepancy in pitch angle. 

\begin{figure*}
% \epsscale{0.6}
\centering
\includegraphics[width=0.6\linewidth]{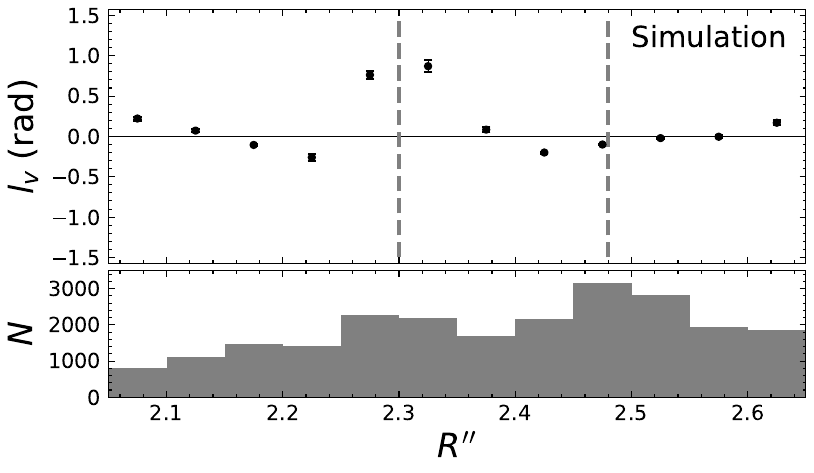}
\caption{Same as Fig.~\ref{fig:R-VD} but obtained from an N-body/hydrodynamic simulations. $R''$ is defined as $\ln R -\phi\times\tan 20^\circ$ to make the direction of $R''$ perpendicular to the shape of the spiral arms (see the snapshot in Fig.~\ref{fig:snap_3581}). The star particles around the inner spiral arm at $R''\sim2.3$ show positive vertex deviations, while those around the outer spiral arm at $R''\sim2.5$ show negative vertex deviations.
  \label{fig:r-lv_2}}
\end{figure*}

Fig.~\ref{fig:r-lv_2} presents vertex deviation as a function of $R''$. Fig.~\ref{fig:r-lv_2} displays $l_v>0$ around a density peak of the inner spiral arm at $R''\sim2.3$, while $l_v<0$ are seen around the outer arm at $R''\sim2.5$. Similar to the observations, the stars in the simulation show the same sign of vertex deviation on either side of these two spiral arms. 
Spiral arms in our numerical simulations follow the dynamic arm scenario, where spiral arms are transient, co-rotating and winding \citep{Dobbs2014}. Hence, the dynamic arm scenario can reproduce two spiral arms with completely opposite kinematic trends, as observed in our sample of Cepheids in the Perseus and Outer arms in the Milky Way.

\subsubsection{Correlation between evolutionary phase of a spiral arm and vertex deviation}\label{sec:dyncorr}

 \begin{figure*}
% \epsscale{0.6}
\centering
\includegraphics[width=\linewidth]{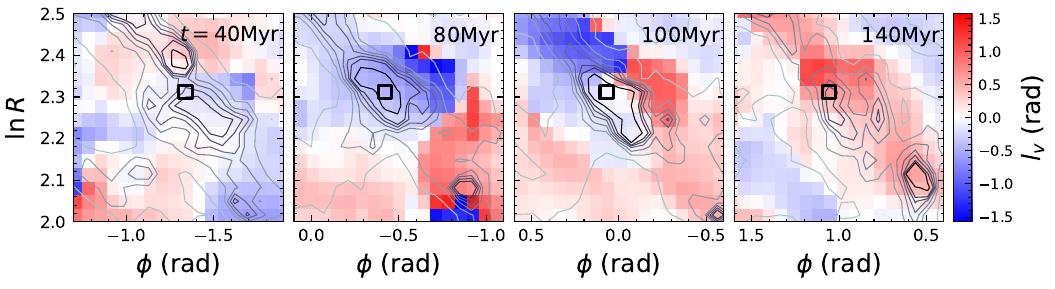}
\caption{Time evolution of the spiral arm and vertex deviation in the N-body/hydrodynamic simulation in the azimuthal angle, $\phi$, and $\ln R$ plane.
Each panel is a snapshot around the single spiral arm focused in Fig.~\ref{fig:t-lv_p}. The colour map shows the vertex deviation of the selected young stars.
The area used to calculate vertex deviation at each time in Fig.~\ref{fig:t-lv_p} is indicated by a black square.
The contour map shows the density distribution of the young star particles, with darker contour lines indicating higher density.
  \label{fig:snap_t-lv}}
\end{figure*}

The dynamic arm scenario predicts different kinematics around spiral arms during growing and disrupting phases of the spiral arms \citep{Baba2013,Grand2014}. We utilised the same N-body/hydrodynamics simulation data used in Section \ref{sec:compneigh} and examined whether there is a correlation between phases of a spiral arm (growing and disrupting) and vertex deviation. For this purpose, 
we selected a conspicuous spiral arm spanning over $5\ltsim R \ltsim15$ kpc. This particular spiral arm was selected due to its clear visibility compared to other spiral arms, helping us to track the evolution of the spiral arm over time. The evolution of vertex deviation is analysed for the time from 3.42 Gyr to 3.60 Gyr in the simulation time, whose time is re-defined to be from $t$=0 to 180 Myr.
Fig.~\ref{fig:snap_t-lv} shows the evolution of the density of this selected arm in contours and the colour map displays the distribution of the vertex deviation.

\begin{figure*}
% \epsscale{0.6}
\centering
\includegraphics[width=0.6\linewidth]{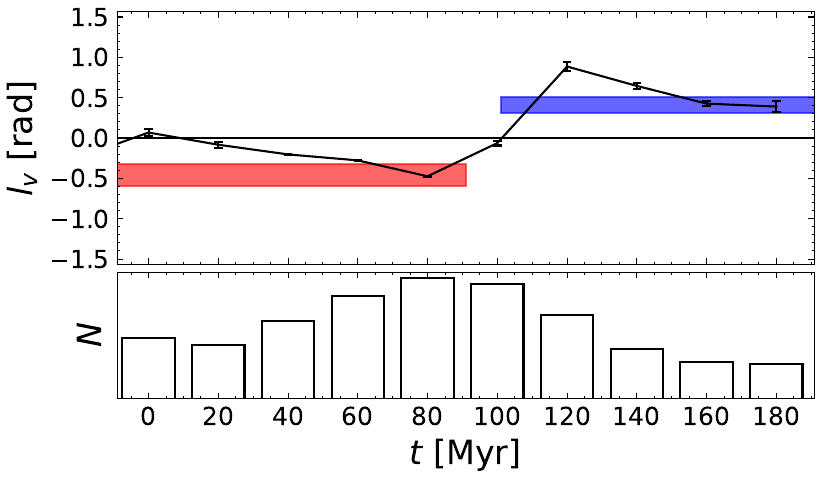}
\caption{Upper: The time evolution of the vertex deviation (the error bars connected with the solid lines)  for the particles within the black square region in Fig.~\ref{fig:snap_t-lv}. 
The purple and red areas indicate the vertex deviations of $l_v=0.40\pm0.13$ and $l_v=-0.56\pm0.15$ obtained from the observations around the Perseus ($2.25<R'<2.35$) and Outer ($2.5<R'<2.6$) arms, respectively. Lower: The evolution of the number of stars in the same selected region around the spiral arm.
When the vertex deviation is negative, as observed in the Cepheids around the Outer arm, the spiral arm is in a growing phase, whereas when it is positive, as in the Perseus arm, the spiral arm is in a disrupting phase.
  \label{fig:t-lv_p}}
\end{figure*}

Fig.~\ref{fig:t-lv_p} shows the vertex deviations computed using particles within the black squares in Fig.~\ref{fig:snap_t-lv}  
around the peak of the spiral arm at a constant radius of $R\sim10$~kpc.
Fig.~\ref{fig:t-lv_p} shows that a spiral arm is growing from $t=0$ to $80$ Myr, i.e. the number of star particles in the black square in Fig.~\ref{fig:snap_t-lv} is increasing in the lower panel, and shows negative vertex deviation. Then, from $t=100$ to $180$ Myr, the spiral arm is disrupting, i.e. the number is decreasing, and positive vertex deviation appears. This suggests that positive and negative vertex deviations are observed in disrupting and growing arms, respectively.

Applying the correlation between evolutionary phases of a spiral arm and the sign of vertex deviation from our simulation to the spiral arms of the Milky Way, we can suggest that the positive vertex deviation around the Perseus arm is due to the Perseus arm being in a disrupting phase.
Meanwhile, the negative vertex deviation around the Outer arm is attributed to the Outer arm being in a growing phase.
Moreover, Fig.~\ref{fig:r-lv_2} shows that both disrupting and growing spiral arms can coexist simultaneously and next to each other in a single galaxy.
Therefore, the observed trend of the Perseus arm and the Outer arm can be explained by different evolutionary phases of the dynamic arm scenario. \\

\begin{figure*}
% \epsscale{0.6}
\centering
\includegraphics[width=0.8\linewidth]{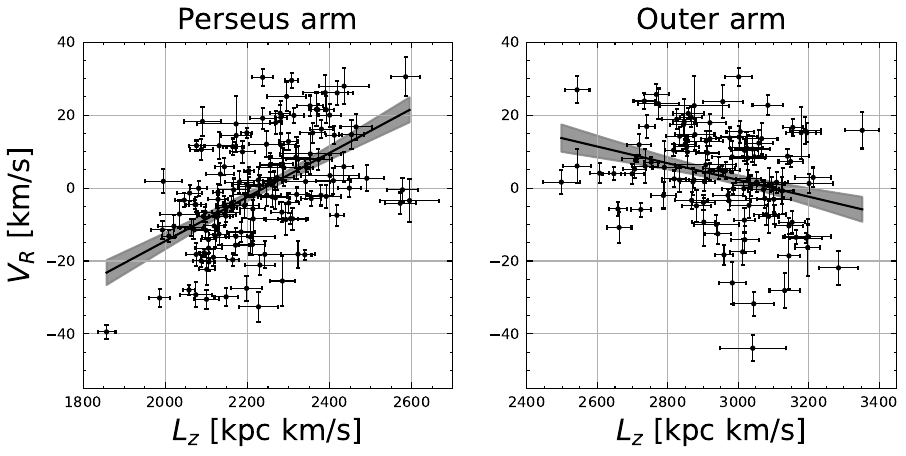}
\caption{\label{fig:lz-vr}
The distribution between $L_z$ and $V_R$ of Cepheids around the Perseus (left) and Outer (right) arms. 
The Cepheids employed in the left panel are within $2.25<R'<2.35$, equivalent to regions (c) and (d) in Fig.~\ref{fig:philnR}. On the other hand, the Cepheids used in the right panel are in the range of $2.5<R'<2.6$, which covers the regions (a) and (b) in Fig.~\ref{fig:philnR}.
The Cepheids around the Perseus arm show a positive correlation between $L_z$ and $V_R$, while the Cepheids around the Outer arm show a negative correlation between $L_z$ and $V_R$. The error bars indicate the uncertainty of $V_R$ and $L_z$ for each Cepheid data. The solid line and the grey shaded region show the results of the linear regression taking into account the uncertainty using \texttt{linmix} package (\url{https://linmix.readthedocs.io/en/latest/}).}
\end{figure*}

The left panel of Fig.~\ref{fig:lz-vr} illustrates that the Cepheids around the Perseus arm typically exhibit larger $V_R$ for those with larger angular momentum, $L_z$. Conversely, the right panel displays that the Cepheids around the Outer arm show smaller $V_R$ for those with larger $L_z$. Stars with lower angular momentum have their guiding centres inside the spiral arm, whereas stars with higher angular momentum have their guiding centres outside the spiral arm.
This provides a view that the Cepheids around the Perseus arm tend to leave their current position and move back towards their guiding centres, because the Cepheids with their guiding centre on the inner side of the spiral arm have inward velocities and the Cepheids with their guiding centre on the outer side of the spiral arm have outward velocities. In other words, the Cepheids around the Perseus arm is going away from each other. This trend is consistent with our suggestion of the disrupting Perseus arm. 
Meanwhile, the Cepheids around the Outer arm tend to move towards each other, because the Cepheids with their guiding centre on the inner side of the spiral arm have outward velocities and the Cepheids with their guiding centre on the outer side of the spiral arm have inward velocities. 
This kinematic trend is also consistent with our picture of the growing Outer arm with the stars gathering toward the Outer arm.

\section{Summary}\label{sec:summary}
In this paper we have reported the first detection of kinematic structure around the Perseus and Outer arms in the Milky Way using Cepheid variable stars. The kinematics of the Cepheids around the spiral arms are not consistent with analytical predictions under the density wave scenarios based on the tight-winding approximation \citep{Roca-Fabrega2014}. On the contrary, in dynamic arm scenarios, neighbouring spiral arms can have different kinematic structures similar to the observed feature. Moreover, comparing their kinematics with numerical simulations, we discovered that the observed different stellar kinematics around the Outer and Perseus arms can be respectively explained by their being growing and disrupting phases of the transient, recurrent and dynamic spiral arms.

Recently, \citet{Asano2023} suggested that the dynamic arm scenario shows a compression of the vertical motion of the star in the growing spiral arm and an expansion in the disrupting spiral arm. They also tentatively found that a hint of vertical expanding motion around the Perseus arm and compressing motion around the Outer arm with \Gaia DR3 data. Interestingly, this is consistent with what we found from the in-plane stellar motion with more precise distance measurements.

Another spiral arm scenario other than the density wave and dynamic arm scenarios is tidally induced spiral arms due to external perturbations \citep{Oh2015,Antoja2022}. In the tidal spiral arm scenario, stars in spiral arms are expected to have $V_R<0$ irrespective of sides of spiral arms \citep{Antoja2022}. In addition, vertex deviation is anticipated to show different signs in front and behind spiral arms, similar to what is expected in the density wave scenario \citep {Antoja2022}.
Unfortunately, neither of these expected features is observed in Cepheids.

It should also be noted that the dynamic arm scenario does not perfectly reproduce our results. The two spiral arms used in Fig.~\ref{fig:r-lv_2} have pitch angles around 20 deg, much larger than the pitch angles of 4 deg observed for the Perseus and Outer arms. In our simulation data, we rarely found spiral arms with pitch angle as small as 4 deg and also covering a large region of the galactic azimuthal angle, like the Perseus and Outer arms. Hence, the dynamic arm scenario still needs to explain the small pitch angles of these two long spiral arms with the different vertex deviation. Furthermore, recent observations by \Gaia inferred the pattern speed of the bar between 33 and 45~$\mathrm{km~s^{-1}}$ kpc \citep{Leung2022,Clarke2022}. This suggests that the bar's Outer Lindblad Resonance (OLR) is expected to be located near the Perseus arm \citep[e.g.][]{Liu2012,Khoperskov2022}. The stellar and gaseous orbits within a bar potential are recognized to vary dependent on the position of resonance \citep[e.g.][]{Dobbs2014}. However, the combination effects of the bar resonances and the dynamic spiral arms are known to be complicated \citep{Hunt2018,Hunt2019}. Our result encourages a further study of the impact of the bar and the dynamic spiral arms on vertex deviation. 

Any successful spiral arm scenario must fully explain our discovery of the diverse trend in stellar kinematics around the spiral arms of the Milky Way. 
The dynamic arm scenario seems to be a promising scenario to explain the observed two opposite kinematic trends of stars around the two spiral arms, the Perseus and Outer arms. This scenario provides clues that these neighbouring spiral arms are in different phases of evolution, i.e. the disrupting Perseus and the growing Outer arm.

\section*{Acknowledgements}
This work is a part of MWGaiaDN, a Horizon Europe Marie Sk\l{}odowska-Curie Actions Doctoral Network funded under grant agreement no. 101072454 and also funded by UK Research and Innovation (EP/X031756/1). This work was also partly supported by the UK's Science \& Technology Facilities Council (STFC grant ST/S000216/1, ST/W001136/1). This work has made use of data from the European Space Agency (ESA) mission {\it Gaia} (\url{https://www.cosmos.esa.int/gaia}), processed by the {\it Gaia}
Data Processing and Analysis Consortium (DPAC, \url{https://www.cosmos.esa.int/web/gaia/dpac/consortium}). Funding for the DPAC has been provided by national institutions, in particular the institutions participating in the {\it Gaia} Multilateral Agreement.
This research was supported by a grant from the Hayakawa Satio Fund awarded by the Astronomical Society of Japan.
JB acknowledges the support of JSPS KAKENHI grant Nos. 21K03633, 22H01259 and 24K07095.

\section*{Data availability}
The datasets used and analysed in this study were derived from data published in the Gaia Archive (\url{https://gea.esac.esa.int/archive}) and from the public classical Cepheid data \citep{Skowron2019} (\url{http://www.astrouw.edu.pl/ogle/ogle4/MILKY_WAY_3D_MAP/Skowron2019_GalClassCephDist.dat}). The simulation data used are available from the corresponding author on a reasonable request.

% \bm{Author contributions}
% N.F. analysed and interpreted the data, and wrote the manuscript. J.B. contributed to sample preparation of the numerical simulation data. N.M., D.K., J.B.,D.T. and M.F. contributed to science discussions.
% N.M. and D.K. helped editing the manuscript.
% All authors have read and approved the manuscript. 

\bibliographystyle{mnras}
\bibliography{example}

%% This command is needed to show the entire author+affiliation list when
%% the collaboration and author truncation commands are used.  It has to
%% go at the end of the manuscript.
%\allauthors

%% Include this line if you are using the \added, \replaced, \deleted
%% commands to see a summary list of all changes at the end of the article.
%\listofchanges

\end{document}